\documentclass[10pt]{article}

\usepackage[english]{babel}
\usepackage{latexsym,amsmath,amssymb,amsbsy,amstext,amscd,amsfonts}
\usepackage{graphics,graphicx}
\usepackage{color}
\usepackage{multirow}
\usepackage{booktabs}
\usepackage{tabularx}

\usepackage{array}
\usepackage{epstopdf}
\usepackage{natbib}
\usepackage{stmaryrd}
\usepackage{cancel}

\newcommand{\beq}{\begin{equation}}
\newcommand{\eeq}{\end{equation}}

\newcommand{\beqar}{\begin{eqnarray}}
\newcommand{\eeqar}{\end{eqnarray}}
\newcommand{\bit}{\begin{itemize}}
\newcommand{\eit}{\end{itemize}}
\newcommand{\benum}{\begin{enumerate}}
\newcommand{\eenum}{\end{enumerate}}
\newcommand{\barr}{\begin{array}}
\newcommand{\earr}{\end{array}}
\def\ds{\displaystyle}

\newcommand{\av}[1]{\langle #1 \rangle}
\newcommand{\jump}[1]{\llbracket #1 \rrbracket}

\def\XXint#1#2#3{{\setbox0=\hbox{$#1{#2#3}{\int}$}
   \vcenter{\hbox{$#2#3$}}\kern-.5\wd0}}

\def\b0{\mbox{\boldmath $0$}}

\def\be{\mbox{\boldmath $e$}}

\def\bn{\mbox{\boldmath $n$}}

\def\bt{\mbox{\boldmath $t$}}
\def\bu{\mbox{\boldmath $u$}}

\newcommand{\bsigma}{\mbox{\boldmath $\sigma$}}

\newcommand{\bvarepsilon}{\mbox{\boldmath $\varepsilon$}}

\def\f0{\ensuremath{\mathbb{O}}}

\newcommand{\mK}{\ensuremath{\mathcal{K}}}

\newcommand{\mT}{\ensuremath{\mathcal{T}}}

\graphicspath{{./}{./figures/}}
\date{\today}

\title{Energy Release Rate, the crack closure integral and admissible singular fields in Fracture Mechanics}

\author{Andrea Piccolroaz$^{(1)}$, Daniel Peck$^{(2)}$, Michal Wrobel$^{(3)}$, Gennady Mishuris$^{(2)}$\footnote{corresponding author}
\\{\it $^{(1)}$\!Department of Civil, Environmental and Mechanical Engineering, University of Trento,}\\
 {\it Trento, Italy}
\\
{\it $^{(2)}$\!Department of Mathematics,
Aberystwyth University, }
\\ {\it Aberystwyth, Wales, United Kingdom}
\\{\it $^{(3)}$\!Department of Civil and Environmental Engineering, University of Cyprus,}\\
 {\it Nicosia, Cyprus}
 }

\begin{document}

\maketitle

\begin{abstract}

One of the assumptions of Linear Elastic Fracture Mechanics is that the crack faces are traction-free or, at most, loaded by bounded tractions. The standard Irwin's crack closure integral, widely used for the computation of the Energy Release Rate, also relies upon this assumption.
However, there are practical situations where the load acting on the crack boundaries is singular. This is the case, for instance, in hydraulic fracturing, where the fluid inside the crack exerts singular tangential tractions at its front.
Another example of unbounded tractions is the case of a rigid line inclusion (anticrack) embedded into an elastic body. In such situations, the classical Irwin's crack closure integral fails to provide the correct value of the Energy Release Rate.
In this paper, we address the effects occurring when square-root singular tractions act at the boundary of a line defect in an elastic solid and provide a generalisation of Irwin's crack closure integral.
The latter yields the correct Energy Release Rate and allows broad applications, including, among others, hydraulic fracturing, soft materials containing stiff inclusions, rigid inclusions, shear bands and cracks characterized by the Gurtin-Murdoch surface stress elasticity.
We present the results in the most general form, where six Stress Intensity Factors are present: three of them are classical SIFs corresponding to the modes I-II-III and computed under the assumption of homogeneous boundary conditions at the defect surfaces, while the other three SIFs are associated with singular admissible tractions (those that lead to a finite ERR value). It is demonstrated that this approach resolves an ambiguity in using the same SIF's terminology in the cases of open cracks and rigid inclusions, among other benefits.

\end{abstract}

\section{Introduction and motivation}\label{Sect:1}

The classical Linear Elastic Fracture Mechanics (LEFM), in the Griffith-Irwin formulation, assumes that the effects of plasticity and dissipative processes are limited to a small crack-tip zone. It also assumes that tractions applied at crack faces are either zero or bounded (if they are singular, the singularity must be weaker than the square-root one).

In \citep{Wrobel2017}, the authors encountered the need to incorporate singular tractions at the crack tip being of the same order as the classic square-root singularity in the study of Hydraulic Fracture (HF).
HF is a phenomenon where a crack in an elastic body is driven by a high-pressure moving fluid (fracking is an example of a technological process, sub-glacial drainage and underground magma flow are examples of natural processes).
The typical approach has been to retain the LEFM framework and to assume that the fluid follows a laminar (Poiseuille) flow. The fluid may also leak-off into the surrounding domain, typically is assumed to follow the Carter Law. The fracture growth is therefore determined by the solid-fluid coupling: i) the tractions on the fracture walls induced by the fluid, ii) the ability of the solid medium to resist this action, both in the direction of crack extension (fracture toughness) and orthogonal to the fluid flow (elastic response). The long-time behaviour of the fracture is typically determined by the rate of fluid loss to leak-off effects.

Due to the strong non-linearity of the solid-fluid interaction, as well as the degeneration of the governing equations at the fracture tip (see \eqref{whyTauSingular}), the modeling of HF began with the use of simplified 1D models: PKN, KGD (plane strain) and radial (penny-shaped). While more advanced models of HF have been developed (reviews can be found in \citep{Adachi2007,Lecampion2018a}), these 1D models continue to form the basis of investigations into the underlying physical processes of HF. For example, the crucial role of the tip asymptotics in solution behaviour was first observed in \citep{Spence} and later expanded upon in \citep{Garagash2011}.

The solvers employing the fracture front asymptotics currently constitute the state-of-the-art approaches; their success has been demonstrated experimentally by \citep{Bunger2008a,JGRB:JGRB51923}. The solvers tend to utilize the level set method first outlined by \citep{Peirce2008}, with a review given by \citep{Peirce2016}, while an open-access general-purpose solver PyFrac was recently released by \citep{ZIA2020107368}. An alternative approach is the development of explicit solvers based on the use of the so-called ``speed equation", a form of Stefan-type boundary condition that tracks the fracture front \citep{KEMP1990,Linkov2011,Linkov2012,MWL2012}, that has been implemented by
\citep{Wrobel2015,Perkowska2015,Peck2018a,Peck2018b}.

All those results have been obtained under the commonly accepted assumption that the shear tractions induced by the fluid on the fracture walls can be neglected. Recently, however, the role of those tractions - previously ignored even though being singular  - has come under increased examination. A preliminary analysis in the framework of the KGD model \citep{Wrobel2017} demonstrated that the shear stress induced by the fluid on the crack surfaces changes the asymptotics near the crack tip.
Naturally, it has a significant effect on the evaluation of the fracture Energy Release Rate (ERR).
It was demonstrated that the differing tip asymptotics offered an alternative physical explanation for the transition in fracture growth behaviour which occurs over time (between viscosity and toughness dominated regimes), while a follow-up investigation demonstrated a possible impact on fracture redirection \citep{Perkowska2017,WPPM2020} and examined the dependence of these effects on fluid rheology \citep{WMP2021}.

It should be noted for the sake of completeness that, although the effects of singular tangential tractions have already attracted some attention \citep{Shen2017,Shen2018,Shen2020,Papanastasiou2018}, the extent to which shear stress on the fracture walls impacts hydraulic fracturing growth remains a controversial issue (see, for example, \citep{Linkov2018,Wrobel2018}). However, this controversy primarily concerns only the role of the tangential tractions on the elasticity equation used within that framework, and does not affect the assessment of the ERR obtained in the original works by \citep{Wrobel2017,Perkowska2017}.

The aforementioned incorporation of fluid shear stresses into planar HF \citep{Wrobel2017,Wrobel2018} required a re-examination of the LEFM due to the square root singularity of the shear traction on the boundary of the fracture at the crack tip, which had previously been ignored. As a result, the asymptotics of the problem needed to be reconsidered, and the corresponding Energy Release Rate (ERR) to be recomputed (following the general principles introduced, independently, by \citep{Cher1967,Cher1968} and \citep{Rice1968}). The most important discrepancy arising from these investigations however was that the ERR computation obtained in \citep{Wrobel2017,Perkowska2017} appears to be in contradiction with that computed in accordance with the standard Irwin's crack closure integral, widely used for the computation of the Energy Release Rate in the framework of the classical LEFM (see Sect.~\ref{Sect:2a}).

This discrepancy demonstrates the lack of a general framework within LEFM for incorporating the effects of square-root singular tractions acting on line-type defects (eg. fractures or rigid inclusions within an otherwise homogeneous material), although some special cases have been considered (see, for example, \citep{Atkinson1995}). The first evaluation of the energy release rate for stiffener growth was given in \citep{Bigoni2008}, further developed in \citep{DalCorso2010,DalCorso2008b} and supported by experimental results on the singular stress state near a stiff lamellar inclusion in \citep{DalCorso2008,Noselli2010} and shear bands \citep{Goudarzi2021}. Another relevant example is given by the crack growth with Gurtin-Murdoch surface stresses on the fracture wall (see \citep{Gorbushin2020} and discussion therein).

Thus, there is a need to extend the framework of LEFM to cases where the square-root singular tractions are also present.
This work aims to establish such a framework in its general setting, extending the classical LEFM. We show that there are six Stress Intensity Factors involved in the analysis: three of them are the classical SIFs corresponding to the modes I-II-III and computed under the assumption of homogeneous boundary conditions at the defect surfaces, while the other three SIFs are associated with singular admissible tractions (those that lead to a finite ERR value).
The analysis involves deriving the Energy Release Rate (ERR) through two separate approaches.

The first is Rice's $J$-integral approach using the formula \eqref{ERR_definition} that leads to the general form of the ERR:
\begin{equation}
\label{eq:generalERR0}
J =
\frac{1}{E_*}
\Big( K_I^2 + K_{II}^2 + (1 + \nu_*)K_{III}^2 \Big) +
\frac{1}{E_*}
\left( K_I \mT_1 - K_{II} \mT_2 - (1 + \nu_*) \mT_{3}^2 \right),
\end{equation}
where $\mT_1, \mT_2, \mT_3$ are three new SIFs associated with the three coefficients in the leading term of the singular tractions $\tau_1, \tau_2, \tau_3$ by $\mT_1 = 2 \sqrt{2\pi}\, \tau_1$, $\mT_2 = 2 \sqrt{2\pi}\, \tau_2$, $\mT_3 = 2 \pi\, \tau_3$.

The second approach is through the principle of virtual work that leads to the generalized Irwin's formula:
\begin{equation}
\label{err_final2a0}
{\cal  G}  = \lim_{\Delta a \to 0} \frac{1}{2 \Delta a} \int_0^{\Delta a}
 \Bigg\{ \av{\sigma_{2i}(a)(X)}  \jump{u_i (a)(X-\Delta a)} - \av{u_{i}(a)(X)} \jump{\sigma_{2i} (a)(X-\Delta a)} \Bigg\}\, dX.
\end{equation}
Details on the notations used in these formulas are given in the main text (see equations
\eqref{eq:generalERR} and \eqref{err_final2a}). However, even from first glance it is clear that these two formulas differ from their classic representations only by the presence of additional terms. We demonstrate that both approaches yield the same, self-consistent, formulation, which reduces to the classical LEFM if the singular tractions vanish ($\tau_1=\tau_2=\tau_3=0$).
Moreover, this resolves the aforemention discrepancy in ERR computations.

This paper is structured as follows.
In Sect.~\ref{Sect:2a} an overview of relevant HF-related results, with tangential tractions induced by the fluid on the crack walls, are given and compared with the expected results from LEFM. Sect.~\ref{Sect:2b} introduces examples from other fields of research where line defects experience square-root tangential tractions, further motivating the work.
In Sect.~\ref{Sect:3}, the general form of the ERR for a crack loaded by (square-root) singular tractions is derived. This is expanded upon in Sect.~\ref{Sect:4}, with the derivation of the general form of the crack closure integral, alongside details of the stress and displacement fields, for any line defect within an elastic material.
In Sect.~\ref{Sect:5} a comparison of the results with existing ones is given, including a comparison with the classical LEFM. Several special cases are also considered: i) the HF with tangential tractions induced by the fluid, ii) the growth/shrinkage of a thin rigid 2D inclusion, iii) the growth of a Mode III crack with singular tractions on its surfaces, iv) the stress intensity factors for a closed crack.
Discussion and concluding remarks are given in Sect.~\ref{Sect:6}.

\section{Preliminary results on singular tractions acting along the linear defects}
\label{Sect:2}
\subsection{Hydraulic Fracture with fluid-induced tangential traction on the crack walls}\label{Sect:2a}



\begin{figure}[htb!]
 \center
 \includegraphics[scale=0.60]{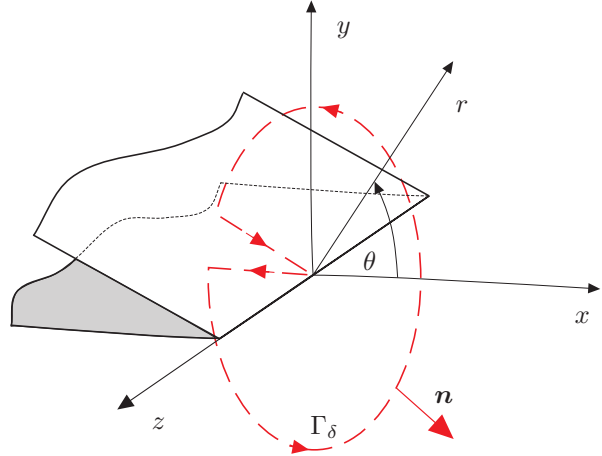}
 \put(-170,10){$z$}
 \put(-10,50){$x$}
 \put(-100,160){$y$}
 \put(-90,70){$\theta$}
 \put(-55,130){$r$}
 \put(-63,20){$\bn$}
 \put(-110,8){$\Gamma_\delta$}
 \caption{A planar crack and its local coordinate system.}
 \label{crack}
\end{figure}

Below we give a short overview of results obtained in the study of hydraulic fracture with shear stress induced upon the crack walls by the fluid, which provides the motivation for this work. In the original paper of \citep{Wrobel2017}, alongside a later extension to the mixed-mode case in \citep{Perkowska2017,WPPM2020}, it was demonstrated that the asymptotics of the displacement and stress fields near the crack-tip ($r\to 0$) are
\begin{equation}
\label{eq:asymu}
\bu(t,r,\theta,z)=\sqrt{\frac{r}{2\pi}}\Big[ K_I(t,z){\bf \Phi}_{I}(\theta)+K_{II}(t,z){\bf\Phi}_{II}(\theta)+K_{III}(t,z){\bf\Phi}_{III}(\theta)+K_f(t,z){\bf \Phi}_{\tau}(\theta)
\Big]+O\left(r \log r\right),
\end{equation}
\begin{equation}
\label{eq:asyms}
\bsigma(t,r,\theta,z)=\frac{1}{\sqrt{2\pi r}}\left[ K_I(t,z){\bf \Psi}_{I}(\theta)+K_{II}(t,z){\bf\Psi}_{II}(\theta)+K_{III}(t,z){\bf\Psi}_{III}(\theta)+K_f(t,z){\bf\Psi}_{\tau}(\theta)\right]
+O\left(\log r\right),
\end{equation}
where $\{r,\theta,z\}$ is a local polar coordinate system linked to the crack tip (see Fig.~\ref{crack}), $K_I$, $K_{II}$ and $K_{III}$ are the classical stress intensity factors (SIFs) associated with homogeneous boundary conditions at the crack faces and $K_f$ is the hydraulic shear stress intensity factor (HSSIF) related to the hydraulic shear stress, $\tau$, induced by the fluid on the crack surfaces  \citep{Wrobel2017}. Respective functions $\bf \Psi_j$, $\bf \Phi_j$ can be found in \citep{Wrobel2017,Perkowska2017}.

As a result, one can obtain
\begin{equation}
\label{eq:asymu_1}
\bu(t,r,0,z)=\frac{1+\nu}{E}
\sqrt{\frac{2r}{\pi}}
\Big\{
(1-2\nu)
\Big\{\,
K_I
\be_r
-
K_{II}
\be_\theta
\Big\}-
2(1-\nu)
K_f
\be_r
\Big\}
+O\left(r \log r\right),
\end{equation}
\begin{equation}
\label{eq:asymu_2}
[\bu]=-
\sqrt{\frac{2r}{\pi}}
\Big\{
\frac{4(1-\nu^2)}{E}
\Big\{\,
\big(K_I+K_{f}\big)
\be_\theta
+
K_{II}
\be_r
\Big\}
-\frac{4(1+\nu)}{E}K_{III}
\be_3
\Big\}
+O\left(r \log r\right),
\end{equation}
\begin{equation}
\label{eq:asyms_1a}
\bsigma_2(t,r,0,z) = \frac{1}{\sqrt{2\pi r}}
\Big\{
\Big(
K_I + 2 (1 - \nu) K_f
\Big) \be_\theta
+
K_{II} \be_r
+
K_{III} \be_3
\Big\}
+ O\left(\log r\right),
\end{equation}
\begin{equation}
\label{eq:asyms_2a}
[\bsigma_2] = -\frac{4(1 - \nu)}{\sqrt{2\pi r}}
K_f \be_r
+ O\left(\log r\right),
\end{equation}
where $[f]=f(\pi)-f(-\pi)$ is the jump of $f$ and $\be_{\{r,\theta,3\}}$ are the unit vectors of the local cylindrical coordinate system.

Note that the hydraulic shear stress intensity factor, $K_f$, is directly related to the mode I SIF, $K_I$, via the factor $\varpi$
\begin{equation}
K_f=\varpi K_I, \quad \varpi=\frac{p_0}{G-p_0}>0,
\label{K_f}
\end{equation}
that, in turn, depends on the coefficient, $p_0$, describing the first term of asymptotics of fluid pressure near the crack tip provided that no lag presents in the process \citep{Wrobel2015}:
\begin{equation}
p(r)\sim p_0\log r, \quad r\to 0.
\label{pressuresing}
\end{equation}
Here, and throughout the paper, $G$ is the elastic shear modulus. For details on how the above is derived, we refer a prospective reader to \citep{Wrobel2017}.

The tangential traction induced on the fracture walls in terms of the crack aperture $w$, fluid velocity $v$ and the normal pressure $p$, follows from lubrication theory
\begin{equation}
\tau(r) = \sigma_{r\theta}(\theta=\pi)=\frac{v(r)}{w(r)}=-\frac{1}{2}w(r)\frac{\partial p}{\partial r}.
\label{whyTauSingular}
\end{equation}
It is clear that this equation will become degenerate at the crack tip ($r\to 0$), as we have the aperture $w\to 0$ while the fluid pressure and it's derivative are singular. The asymptotics of the aperture at the crack tip follows from \eqref{eq:asymu_2}, while those of the pressure derivative are obtained from \eqref{pressuresing}
\[
w \sim w_0 \sqrt{r} , \quad \frac{\partial p}{\partial r} \sim \frac{p_0}{r} , \quad r\to 0.
\]
Thus, the leading term asymptotics of the hydraulic shear stress, i.e. the tangential stress exerted by the fluid on the crack faces, follows immediately as
\begin{equation}
\tau (r) \sim \tau_1\, r^{-1/2}, \quad r \to 0,
\label{tau_shear}
\end{equation}
where the coefficient $\tau_1$ is related to the hydraulic shear stress intensity factor $K_f$ by
\begin{equation}
\tau_1 = \sqrt{\frac{2}{\pi}}\,\, (1 - \nu)\, K_f.
\label{tau_1}
\end{equation}
This asymptotic representation \eqref{tau_shear} -- \eqref{tau_1} can also be obtained from the jump of $\sigma_{r\theta}$.

In the framework of the LEFM, the ERR is computed using the standard $J$-integral arguments \citep{Cher1967,Cher1968,Rice1968}:
\begin{equation}
J = \lim_{ \delta\to0} J^\delta_x(z) = \lim_{ \delta\to0} \int_{\Gamma_\delta} \left\{\frac{1}{2}(\bsigma \cdot \bvarepsilon)n_x-{\bt }_n\cdot \frac{\partial \bf{u}}{\partial x}\right\}ds,
\label{ERR_definition}
\end{equation}
where $\Gamma_\delta$ is a circular contour of radius $\delta$ around the fracture tip, contained in a plane orthogonal to the crack front, $\bf n$ is the outward normal to the contour $\Gamma_\delta$, and ${\bf t}_n = {\boldsymbol \sigma}\bf n$ is the traction vector along $\Gamma_\delta$ (see Fig.~\ref{crack}).

The respective expression for ERR yields (see \citep{Perkowska2017}):
\begin{equation}
\label{err_fracture}
J=
\frac{1 -\nu^2}{E}\left(
K^2_I+K^2_{II}+\frac{1}{1-\nu}K^2_{III}+4(1-\nu)K_IK_f\right),
\end{equation}
which, used as the crack propagation condition, takes the form:
\begin{equation}
\label{err_cond}
J=J_C.
\end{equation}
In a particular case of the Mode I this reduces to:
\begin{equation}
\label{err_final4}
K^2_I+4(1-\nu)K_IK_f=K^2_{IC}\equiv \frac{EJ_C}{1 -\nu^2}.
\end{equation}

The equivalent of the Irwin criterion (typically $K_I=K_{IC}$) can be stated in this description as
\begin{equation}
\label{err_irwin_0}
K_I=\frac{K_{IC}}{\sqrt{1 +4(1-\nu)\varpi}}.
\end{equation}
The right-hand side of the latter can be treated as apparent toughness dependent on the solution at each time step.


On the other hand, one can compute the Energy Release Rate by the classic formula \citep{Irwin1958}, through the so-called crack closure integral or virtual crack closure integral \citep{Rybicki1977,Krueger2004}:
\begin{equation}
\label{err_final1}
{\cal G} = \lim_{\Delta a \to 0} \frac{1}{2 \Delta a} \int_0^{\Delta a}
 \av{\sigma_{2i}(a)}(r) \jump{u_i(a)}(\Delta a - r)\, dr.
\end{equation}
Note that expression \eqref{err_final1} is written in Cartesian coordinates while  \eqref{eq:asymu} -- \eqref{eq:asyms_2a}, used in the computations, are given in the radial coordinates (thus, the vectors $\be_r$ and $\be_\theta$ are differently oriented on the line ahead of the crack tip and the crack surfaces). Here and in what follows, the jumps are computed over the crack surfaces, while the averages on the crack front ahead of the tip.

We refer a prospective reader to the first chapter of a fundamental monograph by \citep{Slepyan2002} for deep and enlightened discussions on basic concepts in fracture mechanics, ERR computations and fracture criteria.

Substituting the asymptotics \eqref{eq:asymu_2} -- \eqref{eq:asyms_1a} into \eqref{err_final1},
we have
\begin{equation}
\label{err_final2}
{\cal G} = \frac{1 -\nu^2}{E}\left(K^2_I+K^2_{II}+\frac{1}{1-\nu}K^2_{III}+(3-2\nu)K_IK_f+2(1-\nu)K_f^2\right).
\end{equation}
Surprisingly, this result is different from \eqref{err_fracture}, seemingly contradicting the classical LEFM.
This was communicated to us by D. Garagash\footnote{private communication}.

This apparent discrepancy requires some explanation, and this paper aims to resolve this central paradox by constructing a general framework within LEFM for any singular field of line-type defects (expanding the results from \citep{Atkinson1973,Atkinson1995}). However, while the original motivation is within HF, this general framework is of importance to numerous other problems. Some of them are outlined in the next subsection, providing both an additional motivation for the construction of this framework as well as important benchmarks against which to examine its conclusions.

\subsection{Singular traction at the defect tip in other physical phenomena}\label{Sect:2b}

The primary motivation for this work is the Hydraulic Fracture; however, similar problems involving singular tractions and computation of the ERR (or the crack closure integral) arise in other areas of research. Two problems of this kind are discussed below.

\subsubsection{Rigid inclusions within biological materials} \label{Sect:2ba}

Consider thin stiff inclusions embedded in a large, soft medium. Such systems are encountered in biomaterials, for example in the study of nanocomposite inclusions within a protein-based bulk material \citep{Goudarzi2021}, or the behaviour of fibrous food during processing such as grinding \citep{Sridhar2013}. In these cases, the inclusions can be modeled as rigid line elements within an infinite elastic isotropic domain.

In the plane strain case, the asymptotic displacement and stress fields near the tip of the inclusion ($r\to0$) can be expressed as \citep{Atkinson1995,Goudarzi2021}
\begin{equation}
\label{rigid}
 \begin{aligned}
\sigma_{xx}(r,\theta) &\sim \frac{\mK_I}{\sqrt{2\pi r}} \cos \frac{\theta}{2} \left\{ 3-2\nu - \sin \frac{\theta}{2} \sin \frac{3\theta}{2} \right\} + \frac{\mK_{II}}{\sqrt{2\pi r}} \sin\frac{\theta}{2} \left\{ 2\nu + \cos\frac{\theta}{2} \cos\frac{3\theta}{2} \right\}, \\
\sigma_{yy} (r,\theta) &\sim -\frac{\mK_I}{\sqrt{2\pi r}} \cos \frac{\theta}{2} \left\{1-2\nu - \sin \frac{\theta}{2} \sin \frac{3\theta}{2} \right\} + \frac{\mK_{II}}{\sqrt{2\pi r}} \sin \frac{\theta}{2} \left\{ 2-2\nu - \cos \frac{\theta}{2} \cos\frac{3\theta}{2} \right\}, \\
\sigma_{xy} (r,\theta) &\sim \frac{\mK_I}{\sqrt{2\pi r}} \sin \frac{\theta}{2} \left\{ 2-2\nu + \cos\frac{\theta}{2} \cos \frac{3\theta}{2} \right\} + \frac{\mK_{II}}{\sqrt{2\pi r}} \cos\frac{\theta}{2} \left\{ 1-2\nu + \sin\frac{\theta}{2} \sin\frac{3\theta}{2} \right\},
\end{aligned}
\end{equation}
where $(x,y)$ and $(r,\theta)$ are cartesian and polar coordinates, respectively, orientated with respect to the inclusion tip as shown in Fig.~\ref{crack}. Meanwhile, $\mK_I,\mK_{II}$ are normalizations of the stress intensity factors $K_I,K_{II}$ of LEFM (see \citep{Noselli2010}). Similar results can also be obtained for the case of multiple ribbon-like inclusions \citep{Atkinson1973}.
When evaluating the asymptotics \eqref{rigid}, the boundary conditions on the inclusion faces are given in terms of displacements, while the resulting tractions acting along the inclusion are:
\[
\sigma_{yy} (r,\pm\pi) \sim \pm \frac{2(1-\nu)}{\sqrt{2\pi r}}\mK_{II},\quad
\sigma_{xy} (r,\pm\pi) \sim \pm \frac{2(1-\nu)}{\sqrt{2\pi r}}\mK_I,\quad r\to0.
\]
Thus, it is not only tangential traction that could be singular but the normal traction as well. Interestingly, these stress components represent the symmetric and skew symmetric parts of solution, respectively.

\subsubsection{Analysis of fractures in nano-structures accounting for surface effects}\label{Sect:2bb}

Another example, where singular traction acts on defect faces, is the Gurtin-Murdoch model that takes the surface stress notion into account \citep{Gurtin1978}. This model considers a body where the behaviour of the surface layer is distinct from that of the underlying elastic material, as is the case for certain class of crystals after cleaving. The stress field at the boundary may therefore include an induced singular traction, depending on the problem geometry. This model has been utilized in the study of elastic domains with coupled surface and internal stresses, for example, in problems involving tangential line loading (e.g. \citep{Le2021}).

Of greater relevance are the models of fracture in elastic material accounting for differences in behaviour of the crack surface and bulk material in Mode I - II \citep{Kim2011a} and Mode III \citep{Kim2010a,Kim2010b}, with applications to the study of various materials. These papers demonstrated that in some cases the surface stress may eliminate the singularity of the stress field at the crack tip. It was even believed that it is a common feature, however, this is not universally true.

In general, incorporating surface stresses results in smoother solutions than in classic elasticity (see, for example, the analysis of weak solutions in \citep{Erem2016,Erem2021}). Thus, one may expect a reduction of any singularities at the defect tip.  However, incorporating surface effects requires a consistent smoothness of the boundary of the domain. Obviously, domains with cracks are not so smooth as required, so some singularities in solutions may still occur.

A more detailed analysis of the stress singularity at the crack tip for a Mode III crack, accounting for Gurtin-Murdoch type surface stress, successfully explained this deviation from the expectations of LEFM \citep{Gorbushin2020}. It was demonstrated that the surface stress effect indeed leads to a square-root singularity of the shear stresses on the surface for the antisymmetric part of the solution and logarithmic singularity for the symmetric one. These results demonstrate that the incorporation of square-root singular tractions into LEFM is still necessary for the analysis of singular field near the defect tips.

\section{General form of the Energy Release Rate for a crack loaded by admissible singular traction}\label{Sect:3}

We start by analysing the Energy Release Rate (ERR) for a crack loaded by singular tractions acting on the crack faces. The assumed singularity of the applied tractions is of the square-root type, $r^{-1/2}$ as $r\to 0$. Note that the analysis performed here is general, as it involves all possible traction components, both in-plane and out-of-plane. Additionally, both symmetric and skew-symmetric loads are considered.

The complete set of singular tractions is illustrated in Table.~\ref{tab:tractions}. Not all traction components are admissible, i.e.\ not all correspond to a finite ERR (non-admissible tractions are associated with unphysical cases of an infinite ERR). Through asymptotic analysis, it can be concluded that the {\it admissible} singular tractions are: symmetric tangential tractions, both in-plane and out-of-plane, and skew-symmetric normal tractions. These tractions produce a stress field with a standard square-root singularity $r^{-1/2}$ at the crack tip. The {\it non-admissible} singular tractions are: skew-symmetric tangential tractions, both in-plane and out-of-plane, and symmetric normal tractions. These tractions produce a stress field with a stronger singularity at the crack tip, namely $r^{-1/2} \log(r)$.

\begin{table}[h!]
 \centering
 \begin{tabular}{@{} l c @{\hspace{1cm}} c @{}}
 \toprule
 Type of tractions & Symmetric & Skew-symmetric \\
 \midrule\noalign{\smallskip}
 \parbox[c]{4.5cm}{In-plane tangential \\ singular tractions} & \parbox[c]{4.5cm}{\centering
 \includegraphics[width=4.5cm]{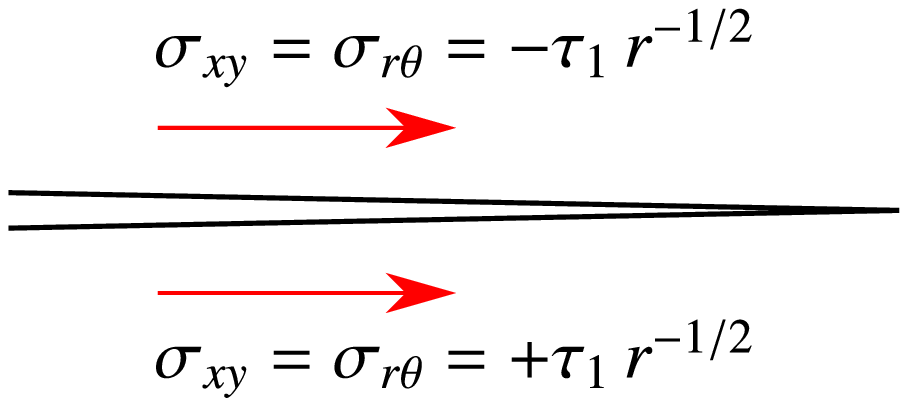} \\ Admissible} & \parbox[c]{4.5cm}{\centering \includegraphics[width=4.5cm]{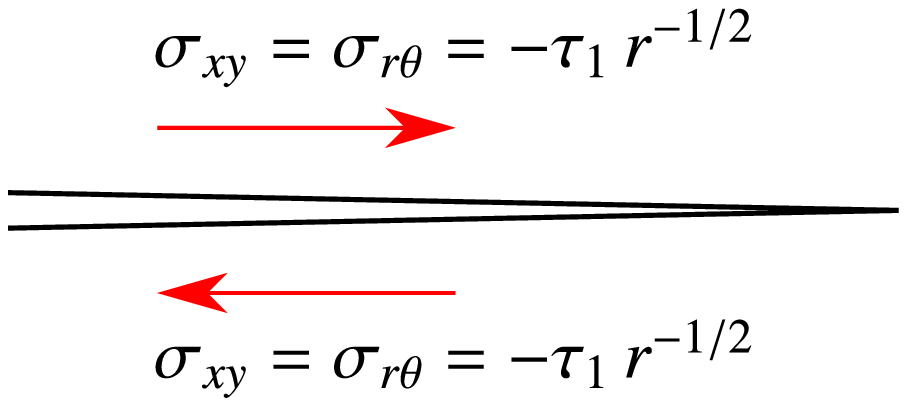} \\ Non-admissible} \\
 \noalign{\smallskip}\midrule\noalign{\smallskip}
 \parbox[c]{4.5cm}{In-plane normal \\ singular tractions} & \parbox[c]{4.5cm}{\centering \includegraphics[width=4.5cm]{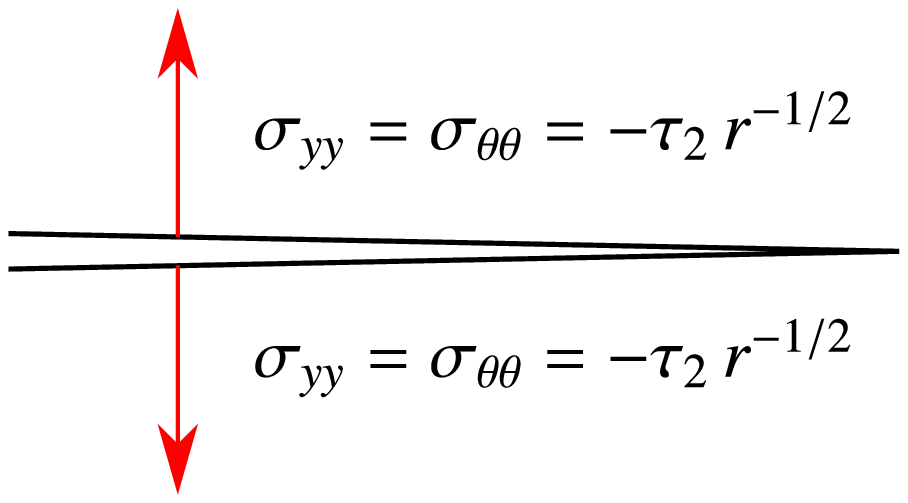} \\ Non-admissible} & \parbox[c]{4.5cm}{\centering \includegraphics[width=4.5cm]{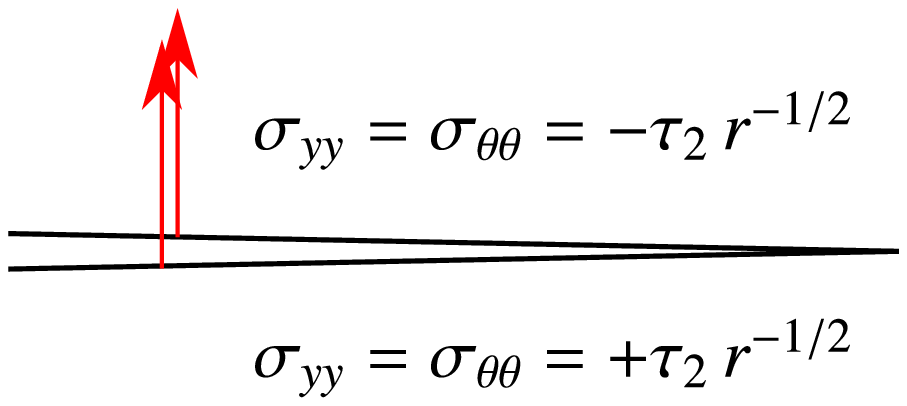} \\ Admissible} \\
 \noalign{\smallskip}\midrule\noalign{\smallskip}
 \parbox[c]{4.5cm}{Out-of-plane tangential \\ singular tractions} &
 \parbox[c]{4.5cm}{\centering \includegraphics[width=4.5cm]{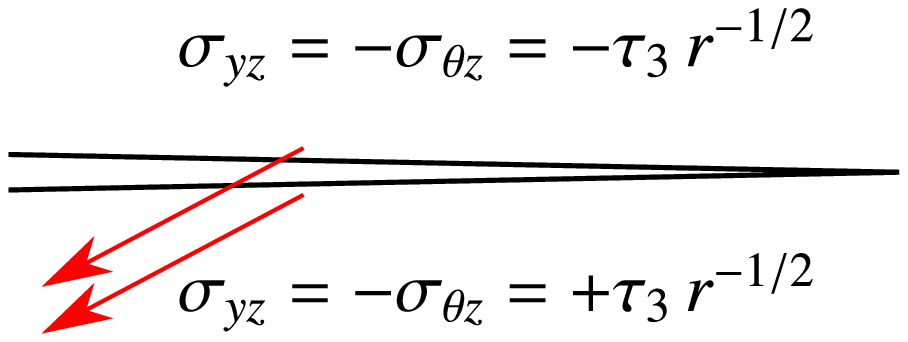} \\ Admissible} & \parbox[c]{4.5cm}{\centering \includegraphics[width=4.5cm]{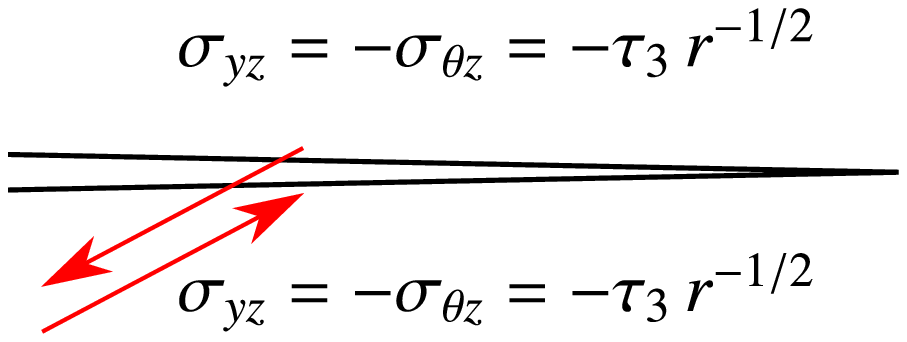} \\ Non-admissible} \\
 \noalign{\smallskip}\bottomrule
 \end{tabular}
 \caption[]{Admissible and non-admissible singular tractions applied to the crack faces. }
 \label{tab:tractions}
\end{table}

{\bf Remark:} For the purpose of this investigation, an admissible traction is one which produces a finite ERR. The justification for this is as follows. In the case when a non-admissible load is applied, the crack propagation is dynamically unstable (as the ERR is infinite). Thus, it will inevitably outrun the source of any predefined singular load acting at the crack tip, and the classical LEFM instead governs the propagation starting from some moment. Conversely, for the continuous application of a singular tangential traction at the crack tip, one requires that the energy induced into the fracture be finite. Therefore, a finite ERR is required (the traction is admissible). Thus, only cases with a finite ERR need be considered in the development of this framework.

The case of in-plane tangential singular traction was analysed by \citep{Wrobel2017}, see Sect.~\ref{Sect:2a}. We now report the full asymptotic solution, including the ERR for the general case in which all admissible singular tractions act on the crack faces.

The assumed singular tractions acting on the crack faces ($\theta = \pm \pi$) have the following crack-tip asymptotics:
\begin{equation}
\label{stress_boundary}
\begin{aligned}
 \sigma_{xy}(\theta &= \pm\pi) = \sigma_{r\theta}(\theta = \pm\pi) = \mp \tau_1 r^{-1/2}+O\big(r^{-1/2+\varepsilon}\big), \\
 \sigma_{yy}(\theta &= \pm\pi) = \sigma_{\theta\theta}(\theta = \pm\pi) = \mp \tau_2 r^{-1/2}+O\big(r^{-1/2+\varepsilon}\big), \quad  r\to0,\\
 \sigma_{yz}(\theta &= \pm\pi) = -\sigma_{\theta z}(\theta = \pm\pi) = \mp \tau_3 r^{-1/2}+O\big(r^{-1/2+\varepsilon}\big),
\end{aligned}
\end{equation}
where $\varepsilon$ is a positive constant.

Here the three values $\tau_1$, $\tau_2$ and $\tau_3$ are the multipliers of the leading singular terms of the admissible tractions that give rise to three new SIFs as demonstrated below. These SIFs, denoted in this work as $\mT_1$, $\mT_2$ and $\mT_3$, are local in the sense that they fully define the asymptotic solution near the crack tip, in contrast with the three classic SIFs ($K_I$, $K_{II}$ and $K_{III}$) that should be found from the global solution. However, this difference is not as straightforward as one might assume given the previous arguments. It is crucial to note that the local SIFs may depend on the global ones (see, for example, \eqref{K_f} and \eqref{tau_1}).

The corresponding asymptotics of displacement and stress fields near the crack-tip are:


%

\paragraph{Plane strain / Plane stress:}


\begin{multline}
\label{eq:asymu1}
\begin{bmatrix}
u_r \\
u_\theta
\end{bmatrix}
= \frac{1}{E_*} \sqrt{\frac{r}{2 \pi}}
\Bigg\{
K_{I}
\begin{bmatrix}
\ds \cos \frac{\theta}{2} \left( 3 - \nu_* - (1 + \nu_*) \cos \theta \right) \\[3mm]
\ds -\sin \frac{\theta}{2} \left( 3 - \nu_* - (1 + \nu_*) \cos \theta \right)
\end{bmatrix}
+ K_{II}
\begin{bmatrix}
\ds -\sin \frac{\theta}{2} \left( 1 - 3\nu_* - 3(1 + \nu_*) \cos \theta \right) \\[3mm]
\ds -\cos \frac{\theta}{2} \left( 5 + \nu_* - 3(1 + \nu_*) \cos \theta \right)
\end{bmatrix} \\
+ \tau_{1}
\begin{bmatrix}
\ds - 2 \sqrt{2\pi} (1 + \nu_*) \cos \frac{3\theta}{2}  \\[3mm]
\ds 2 \sqrt{2\pi} (1 + \nu_*) \sin \frac{3\theta}{2}
\end{bmatrix}
+ \tau_{2}
\begin{bmatrix}
\ds - 2 \sqrt{2\pi} (1 + \nu_*) \sin \frac{3\theta}{2}  \\[3mm]
\ds - 2 \sqrt{2\pi} (1 + \nu_*) \cos \frac{3\theta}{2}
\end{bmatrix}
\Bigg\} +O\big(r^{1/2+\varepsilon}\big), \quad  r\to0,
\end{multline}

\begin{multline}
\label{eq:asyms1}
\begin{bmatrix}
\sigma_{rr} \\
\sigma_{\theta\theta} \\
\sigma_{r\theta}
\end{bmatrix}
= \frac{1}{\sqrt{2\pi r}}
\Bigg\{
K_{I}
\begin{bmatrix}
\ds \frac{5}{4} \cos \frac{\theta}{2} - \frac{1}{4} \cos \frac{3 \theta}{2} \\[3mm]
\ds \cos^3 \frac{\theta}{2} \\[3mm]
\ds \frac{1}{4} \csc \frac{\theta}{2} \sin^2 \theta
\end{bmatrix}
+ K_{II}
\begin{bmatrix}
\ds -\frac{5}{4} \sin \frac{\theta}{2} + \frac{3}{4} \sin \frac{3 \theta}{2} \\[3mm]
\ds -\frac{3}{4} \csc \frac{\theta}{2} \sin^2 \theta \\[3mm]
\ds \frac{1}{4} \cos \frac{\theta}{2} + \frac{3}{4} \cos \frac{3 \theta}{2}
\end{bmatrix} \\
+ \tau_{1}
\begin{bmatrix}
\ds - \sqrt{2\pi} \cos \frac{3\theta}{2} \\[3mm]
\ds \sqrt{2\pi} \cos \frac{3\theta}{2} \\[3mm]
\ds \sqrt{2\pi} \sin \frac{3\theta}{2}
\end{bmatrix}
+ \tau_{2}
\begin{bmatrix}
\ds - \sqrt{2\pi} \sin \frac{3\theta}{2}  \\[3mm]
\ds \sqrt{2\pi} \sin \frac{3\theta}{2} \\[3mm]
\ds -\sqrt{2\pi} \cos \frac{3\theta}{2}
\end{bmatrix}
\Bigg\}+O\big(r^{-1/2+\varepsilon}\big), \quad  r\to0,
\end{multline}
where
\[
E_* = \frac{E}{1 - \nu^2}, \quad \nu_* = \frac{\nu}{1 - \nu}, \quad \sigma_{zz} = \nu (\sigma_{rr} + \sigma_{\theta\theta}),\quad
\varepsilon_{zz}=0,
 \]
for plane strain, and
\[
E_* = E, \quad \nu_* = \nu, \quad \sigma_{zz} = 0,\quad \varepsilon_{zz}=-\frac{\nu}{1-\nu}\big(\varepsilon_{rr} + \varepsilon_{\theta\theta}\big),
\]
for plane stress. Note that the plane strain framework corresponds to the general 3D plane crack (planar 3D crack in the HF terminology).

\paragraph{Antiplane shear:}

\begin{equation} \label{eq:asymu1aaa}
u_z = 4 \frac{1 + \nu}{E} \sqrt{\frac{r}{2 \pi}}
\Bigg\{
K_{III} \sin \frac{\theta}{2} - \tau_3 \sqrt{2\pi} \cos \frac{\theta}{2}
\Bigg\} +O\big(r^{1/2+\varepsilon}\big), \quad  r\to0,
\end{equation}

\begin{equation} \label{eq:asyms1aaa}
\begin{bmatrix}
\sigma_{rz} \\
\sigma_{\theta z}
\end{bmatrix}
= \frac{1}{\sqrt{2\pi r}}
\Bigg\{
K_{III}
\begin{bmatrix}
\ds \sin \frac{\theta}{2} \\[3mm]
\ds \cos \frac{\theta}{2}
\end{bmatrix}
+ \tau_3
\begin{bmatrix}
\ds -\sqrt{2\pi} \cos \frac{\theta}{2} \\[3mm]
\ds \sqrt{2\pi} \sin \frac{\theta}{2}
\end{bmatrix}
\Bigg\}+O\big(r^{-1/2+\varepsilon}\big), \quad  r\to0.
\end{equation}

For example, it follows from \eqref{eq:asymu1}, \eqref{eq:asymu1aaa} that the displacements on the crack surfaces are:
\begin{equation}
\label{displacement_boundary}
\begin{aligned}
 u_{x}(\theta &= \pm\pi) = - u_{r}(\theta = \pm\pi) = \pm \sqrt{\frac{r}{2\pi}} \frac{4}{E_*} \left( K_{II} - \sqrt{\frac{\pi}{2}} (1 + \nu_*) \tau_2 \right) +O\big(r^{1/2+\varepsilon}\big), \\
 u_{y}(\theta &= \pm\pi) = - u_{\theta}(\theta = \pm\pi) = \pm \sqrt{\frac{r}{2\pi}} \frac{4}{E_*} \left( K_{I} + \sqrt{\frac{\pi}{2}} (1 + \nu_*) \tau_1 \right)+O\big(r^{1/2+\varepsilon}\big),\\
 u_{z}(\theta &= \pm\pi) = \pm \sqrt{\frac{r}{2\pi}}\frac{4(1+\nu)}{E}K_{III}+O\big(r^{1/2+\varepsilon}\big),
\end{aligned}
\quad  r\to0.
\end{equation}

Using the formula \eqref{ERR_definition}, the general form of the ERR is obtained as follows
\begin{equation}
\label{eq:generalERR}
J = \frac{1}{E_*} \left( K_I^2 + K_{II}^2 +  2\sqrt{2\pi} K_I \tau_1 - 2\sqrt{2\pi} K_{II} \tau_2 \right)+ \frac{1 + \nu_*}{E_*} \Big(K_{III}^2- 2\pi \tau_{3}^2\Big).
\end{equation}
Introducing now three new SIFs as $\mT_1 = 2\sqrt{2\pi}\, \tau_1$, $\mT_2 = 2\sqrt{2\pi}\, \tau_2$ and $\mT_3 = 2\pi\, \tau_3$ one obtains the generalised ERR formula \eqref{eq:generalERR0} given in the introduction.

Note that $(1+\nu)/E=(1+\nu_*)/E_*$. Thus, this representation is meaningful for the plain strain, plain stress, Mode III case and in the general 3D case at each point along the smooth crack front in the local coordinate system.

Combining \eqref{stress_boundary} and \eqref{displacement_boundary},
one can model any boundary condition on the sides of a linear defect and reconstruct the respective
singular field that yields a finite ERR for the problem, together with the value of the ERR computed from the first principles in \eqref{eq:generalERR}.
This general result is consistent with the one that has been obtained by the same technique (Rice's $J$-integral approach) previously in \citep{Wrobel2017}, and thus, the aforementioned mismatch between the ERR values computed by formulas \eqref{ERR_definition} and the crack closure integral \eqref{err_final1} is still to be resolved (see Sect.~\ref{Sect:2a}).

\section{General form of the crack closure integral}\label{Sect:4}

\begin{figure}[!htb]
 \centering
 \includegraphics[scale=0.60]{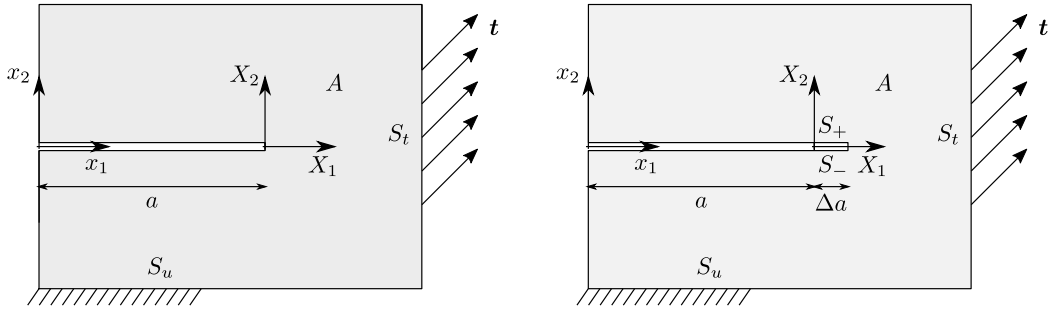}
 \caption{Two linear elastic bodies containing a crack} 
 \label{fig}
\end{figure}

We revisit the classic derivation of the crack closure integral that takes into account an arbitrary admissible singular field near the defect tip.  Following the standard approach (see for example \citep{Krueger2004}), let us consider two linear elastic bodies that are otherwise identical except that the first one has a crack of length $a$ whereas the second has a crack of length $a + \Delta a$, see Fig.~\ref{fig}.

The potential energy per unit thickness of the first body in the absence of body forces is
\begin{equation}
 \Pi(a) = \int_A W(a)\, dA - \int_{S_t} t_i(a) u_i (a)\, ds,
\end{equation}
where $A$ is the domain occupied by the body, $S_t$ is the portion of the boundary on which the tractions $\bt$
are prescribed and $W$ is the strain energy density.
Similarly, for the second body the potential energy is
\begin{equation}
 \Pi(a + \Delta a) = \int_A W(a + \Delta a)\, dA - \int_{S_t'} t_i(a + \Delta a) u_i (a + \Delta a)\, ds,
\end{equation}
where $S_t'$ is the union of $S_t$ and the additional crack surfaces associated with the increment of crack length $\Delta a$,
\begin{equation}
S_t' = S_t \cup S_+ \cup S_-\, .
\end{equation}
We assume that both bodies are clamped along the surface $S_u$:
\begin{equation}
u_i(a)=u_i(a+\Delta a)=0, \quad s\in S_u.
\label{fixed}
\end{equation}
Applying Clapeyron's theorem we obtain
\begin{equation}
 \Pi(a) = - \frac{1}{2} \int_{S_t} t_i(a) u_i (a)\, ds,
\end{equation}
\begin{equation}
 \Pi(a + \Delta a) = - \frac{1}{2} \int_{S_t} t_i (a + \Delta a) u_i (a + \Delta a)\, ds - \frac{1}{2} \int_{S_+ \cup S_-} t_i (a + \Delta a) u_i (a + \Delta a)\, ds.
\end{equation}
The change of the potential energy $ \Delta \Pi = \Pi(a + \Delta a) - \Pi(a)$ can be computed as
\begin{equation}
 \Delta \Pi =
 - \frac{1}{2} \int_{S_t} \Big( t_i(a + \Delta a) u_i (a + \Delta a) - t_i(a) u_i(a) \Big)\, ds
 - \frac{1}{2} \int_{S_+ \cup S_-} t_i(a + \Delta a) u_i(a + \Delta a)\, ds.
 \label{delta}
\end{equation}

Due to \eqref{fixed}, the first integral in \eqref{delta} can be extended over the entire boundary $S = S_t \cup S_u$ of the first body:
\begin{equation}
\label{delta1}
 \Delta \Pi =
 - \frac{1}{2} \int_{S} \Big( t_i(a + \Delta a) u_i(a + \Delta a) - t_i(a) u_i(a) \Big)\, ds
 - \frac{1}{2} \int_{S_+ \cup S_-} t_i(a + \Delta a) u_i(a + \Delta a)\, ds.
\end{equation}
Expanding the last integral, we obtain
\begin{multline}
\label{delta2}
 \Delta \Pi =
 - \frac{1}{2} \int_{S} \Big( t_i(a + \Delta a) u_i(a + \Delta a) - t_i(a) u_i(a) \Big)\, ds \\
 + \frac{1}{2} \int_{a}^{a + \Delta a} \sigma_{2i}^+(a + \Delta a) u_i^+(a + \Delta a)\, dx_1
 - \frac{1}{2} \int_{a}^{a + \Delta a} \sigma_{2i}^-(a + \Delta a) u_i^-(a + \Delta a)\, dx_1.
\end{multline}
Betti's reciprocal theorem implies that
\begin{equation}
\label{betti}
 \int_A \sigma_{ij}(a) \varepsilon_{ij}(a + \Delta a)\, dA =
 \int_A \sigma_{ij}(a + \Delta a) \varepsilon_{ij}(a)\, dA.
\end{equation}

The principle of virtual work applied separately to the two integrals appearing in \eqref{betti} yields
\begin{multline}
\label{work1}
 \int_A \sigma_{ij}(a) \varepsilon_{ij}(a + \Delta a)\, dA =
 \int_{S} t_i(a) u_i(a + \Delta a)\, ds - \int_{a}^{a + \Delta a} \sigma_{2i}^+(a) u_i^+(a + \Delta a)\, dx_1 \\
 + \int_{a}^{a + \Delta a} \sigma_{2i}^-(a) u_i^-(a + \Delta a)\, dx_1,
\end{multline}
and
\begin{multline}
\label{work2}
 \int_A \sigma_{ij}(a + \Delta a) \varepsilon_{ij}(a)\, dA =
 \int_{S} t_i(a + \Delta a)  u_i(a)\, ds - \int_{a}^{a + \Delta a} \sigma_{2i}^+(a + \Delta a) u_i^+(a)\, dx_1 \\
 + \int_{a}^{a + \Delta a} \sigma_{2i}^-(a + \Delta a) u_i^-(a)\, dx_1,
\end{multline}
As a result, one concludes that
\begin{multline}
\label{work3}
 \int_{S} t_i(a) u_i (a + \Delta a)\, ds - \int_{S} t_i(a + \Delta a)  u_i (a)\, ds
 - \int_{a}^{a + \Delta a} \sigma_{2i}^+(a) u_i^+ (a + \Delta a)\, dx_1 + \int_{a}^{a + \Delta a} \sigma_{2i}^-(a) u_i^- (a + \Delta a)\, dx_1 \\[2mm]
+\int_{a}^{a + \Delta a} \sigma_{2i}^+(a + \Delta a) u_i^+ (a)\, dx_1
 - \int_{a}^{a + \Delta a} \sigma_{2i}^-(a + \Delta a) u_i^- (a)\, dx_1=0,
\end{multline}
and thus the energy increment \eqref{delta2} can be reformulated as:
\begin{multline}
\label{delta3}
 \Delta \Pi =
 - \frac{1}{2} \int_{S} \Big( t_i(a + \Delta a) u_i(a + \Delta a) - t_i(a) u_i(a) + t_i(a + \Delta a) u_i(a) - t_i(a) u_i(a + \Delta a) \Big)\, ds
 \\
 + \frac{1}{2} \int_{a}^{a + \Delta a} \sigma_{2i}^+(a + \Delta a) u_i^+ (a + \Delta a)\, dx_1
 - \frac{1}{2} \int_{a}^{a + \Delta a} \sigma_{2i}^-(a + \Delta a) u_i^- (a + \Delta a)\, dx_1\\
-\frac{1}{2}\int_{a}^{a + \Delta a} \sigma_{2i}^+(a) u_i^+ (a + \Delta a)\, dx_1 + \frac{1}{2}\int_{a}^{a + \Delta a} \sigma_{2i}^-(a) u_i^- (a + \Delta a)\, dx_1 \\[2mm]
+ \frac{1}{2}\int_{a}^{a + \Delta a} \sigma_{2i}^+(a + \Delta a) u_i^+ (a)\, dx_1
 -\frac{1}{2}\int_{a}^{a + \Delta a} \sigma_{2i}^-(a + \Delta a) u_i^- (a)\, dx_1.
\end{multline}
Let's now assume for simplicity that:
\begin{itemize}
\item $t_i(a)$ and $t_i(a+\Delta a)$ coincide on that part of the external boundary, $S_e$, that does not contain the crack surfaces $S_0=S_0^+ \cup S_0^-$, ($S_e = S\, \backslash S_0$);
\item and $S_u\cap S_0={\O}$.
\end{itemize}

Note that these assumptions concerning the external boundary $S_e$ can be replaced by the weaker assumption that both solutions are smooth along the external boundary. This condition is also assumed and crucial for the derivation of the classical LEFM, although it is rarely mentioned.

\vspace{2mm}

As a result, the first integral in \eqref{delta3} can be computed over surface $S_0$ instead of $S$.
Now, the continuity of the physical fields ahead of the crack tip ($x_1\in(a,a+\Delta a)$),
\begin{equation}
 \sigma_{2i}^+(a) = \sigma_{2i}^-(a) = \langle \sigma_{2i}(a) \rangle,\quad
 u_{i}^+(a) = u_{i}^-(a) = \langle u_{i}(a) \rangle,
\label{cont}
\end{equation}
yields:
\begin{multline}
\label{delta4a}
 \Delta \Pi =
 - \frac{1}{2} \int_{S_0} \Big( t_i(a + \Delta a) u_i(a + \Delta a) - t_i(a) u_i(a) + t_i(a + \Delta a) u_i(a) - t_i(a) u_i(a + \Delta a) \Big)\, ds \\
 + \frac{1}{2} \int_{a}^{a + \Delta a} \jump{\sigma_{2i}(a + \Delta a) u_i(a + \Delta a)}\, dx_1 \\
 - \frac{1}{2} \int_{a}^{a + \Delta a} \av{\sigma_{2i}(a)} \jump{u_i (a + \Delta a)}\, dx_1
 + \frac{1}{2} \int_{a}^{a + \Delta a} \jump{\sigma_{2i}(a + \Delta a)} \av{u_{i}(a)}\, dx_1,
\end{multline}
where
\begin{equation}
\jump{f(a+\Delta a)} = f^+(a+\Delta a) - f^-(a+\Delta a).
\label{jump}
\end{equation}
Note that:
\begin{equation}
\label{err}
{\cal G} = - \frac{\partial \Pi}{\partial a} = \lim_{\Delta a \to 0} \frac{1}{2 \Delta a} \int_a^{a+\Delta a}
 \Bigg\{ \av{\sigma_{2i}(a)} \jump{u_i(a + \Delta a)} - \av{u_{i}(a)} \jump{\sigma_{2i}(a + \Delta a)} \Bigg\}\, dx_1 + {\cal G}_1^* + {\cal G}_2^*,
\end{equation}
where
\begin{equation}
\label{err_1_*}
{\cal G}_1^* = - \lim_{\Delta a \to 0} \frac{1}{2 \Delta a} \int_{a}^{a + \Delta a} \jump{\sigma_{2i}(a + \Delta a) u_i(a + \Delta a)}\, dx_1,
\end{equation}
\begin{equation}
\label{err_2_*}
{\cal G}_2^* = \lim_{\Delta a \to 0} \frac{1}{2 \Delta a} \int_{S_0} \Big( t_i(a + \Delta a) u_i(a + \Delta a) - t_i(a) u_i(a) + t_i(a + \Delta a) u_i(a) - t_i(a) u_i(a + \Delta a) \Big)\, ds.
\end{equation}
Taking into account estimates
\eqref{stress_boundary} and
\eqref{displacement_boundary}, the leading terms cancel one another out and
the integral in \eqref{err_1_*} behaves as $O(\Delta a^{1+\varepsilon})$ when $\Delta a\to0$. Thus:
\[
{\cal G}_1^* = 0.
\]

In order to determine the term, ${\cal G}_2^*$, in \eqref{err}, we split it over two subdomains: the first being a small region near the crack tip and the second being the remaining domain. The first can be estimated in the same way as it was done when considering ${\cal G}_1^*$. The second integral over the domain \eqref{err_2_*} vanishes if one assumes smoothness (regularity) of both $u_i$ and $t_i$ along that domain. As a result
\[
 {\cal G}_2^* =0,
\]
since both limits are zeros, as is the case for classical LEFM.

Summarising, we have found that:
\begin{equation}
\label{err_final1b}
{\cal  G} = \lim_{\Delta a \to 0} \frac{1}{2 \Delta a} \int_a^{a+\Delta a}
 \Bigg\{ \av{\sigma_{2i}(a)}  \jump{u_i (a + \Delta a)} - \av{u_{i}(a)} \jump{\sigma_{2i} (a + \Delta a)} \Bigg\}\, dx_1.
\end{equation}
Next, observing as in the classic approach that locally ($a < x_1 < a + \Delta a$), $\Delta a \to 0+$
\begin{equation}
\label{e0}
\sigma_{2i} (a+\Delta a)(x_1)=\sigma_{2i} (a)(x_1-\Delta a),\quad u_{i} (a+\Delta a)(x_1)=u_{i} (a)(x_1- \Delta a),
\end{equation}
and changing the coordinate system: $X=x_1-a$, we finally arrive at a generalised form of the classic virtual crack closure integral \eqref{err_final1}
\begin{equation}
\label{err_final2a}
{\cal  G}  = \lim_{\Delta a \to 0} \frac{1}{2 \Delta a} \int_0^{\Delta a}
 \Bigg\{ \av{\sigma_{2i}(a)(X)}  \jump{u_i (a)(X-\Delta a)} - \av{u_{i}(a)(X)} \jump{\sigma_{2i} (a)(X-\Delta a)} \Bigg\}\, dX.
\end{equation}

The crack closure integral can be written also as a convolution integral by using a polar coordinate system attached to the crack tip. In this case we have
\begin{equation}
\label{e1}
\sigma_{\theta j}(a + \Delta a)(r) = \sigma_{\theta j}(a)(\Delta a - r), \quad u_{j}(a + \Delta a)(r) = u_{j}(a)(\Delta a - r),\quad j=r,\theta,z.
\end{equation}
The formula for the energy release rate becomes
\begin{equation}
\label{err_final3a}
\begin{aligned}
{\cal  G} =
\lim_{\Delta a \to 0} \frac{1}{2 \Delta a} \int_0^{\Delta a} \Bigg\{
& \av{\sigma_{\theta r}(a)(r)} \jump{-u_r(a)(\Delta a - r)} - \av{u_{r}(a)(r)} \jump{\sigma_{\theta r}(a)(\Delta a - r)} +\\
& \av{\sigma_{\theta \theta}(a)(r)} \jump{-u_\theta(a)(\Delta a - r)} - \av{u_{\theta}(a)(r)} \jump{\sigma_{\theta \theta}(a)(\Delta a - r)}+ \\
& \av{\sigma_{\theta z}(a)(r)} \jump{u_z(a)(\Delta a - r)} - \av{u_{z}(a)(r)} \jump{-\sigma_{\theta z}(a)(\Delta a - r)}
\Bigg\}\, dr.
\end{aligned}
\end{equation}

Note that both generalised forms of the virtual crack closure integral \eqref{err_final2a} and \eqref{err_final3a} coincide with the classic one \eqref{err_final1}
provided that the standard LEFM boundary conditions hold along the crack faces, in which case the second part of the new integral \eqref{err_final2a} vanishes. In general, the second part of the integral \eqref{err_final2a} represents the energy spent on deformation of the material in front of the crack, while the first part describes the amount of energy consumed during the crack opening (as in the classic case).

The jumps and the average values of stress and displacement
appeared in the crack closure integrals \eqref{err_final2a} or \eqref{err_final3a}
can be estimated near the crack tip in the following way ($r\to0$):
\begin{equation}
\begin{aligned}\label{Sigma_Tang1}
\av{\sigma_{12}} (r) &= \av{\sigma_{r\theta}} (r) \sim \frac{K_{II} - \sqrt{2\pi}\, \tau_2}{\sqrt{2\pi r}}, \quad
& \jump{\sigma_{12}} (r) = \jump{\sigma_{r\theta}} (r) \sim -\frac{2\tau_1}{\sqrt{r}}, \\
\av{\sigma_{22}} (r) &= \av{\sigma_{\theta\theta}} (r) \sim \frac{K_{I} + \sqrt{2\pi}\, \tau_1}{\sqrt{2\pi r}}, \quad
& \jump{\sigma_{22}} (r) = \jump{\sigma_{\theta\theta}} (r) \sim -\frac{2\tau_2}{\sqrt{r}}, \\
\av{\sigma_{23}} (r) &= \av{\sigma_{\theta 3}} (r) \sim \frac{K_{III}}{\sqrt{2\pi r}}, \quad
& \jump{\sigma_{23}} (r) = -\jump{\sigma_{\theta 3}} (r) \sim -\frac{2\tau_{3}}{\sqrt{r}},
\end{aligned}
\end{equation}
\begin{equation}
\begin{aligned} \label{Disp_Tang1}
\av{u_{1}} (r) &= \av{u_{r}} (r) \sim \sqrt{\frac{2r}{\pi}} \frac{1}{E_*} \left[ (1 - \nu_*) K_I - \sqrt{2\pi}\, (1 + \nu_*) \tau_1 \right], \\
\jump{u_{1}} (r) &= -\jump{u_{r}} (r) \sim \sqrt{\frac{2r}{\pi}} \frac{4}{E_*} \left[ K_{II} - \sqrt{\frac{\pi}{2}}\, (1 + \nu_*) \tau_2 \right], \\
\av{u_{2}} (r) &= \av{u_{\theta}} (r) \sim -\sqrt{\frac{2r}{\pi}} \frac{1}{E_*} \left[ (1 - \nu_*) K_{II} + \sqrt{2\pi}\, (1 + \nu_*) \tau_2 \right], \\
\jump{u_{2}} (r) &= -\jump{u_{\theta}} (r) \sim \sqrt{\frac{2r}{\pi}} \frac{4}{E_*} \left[ K_{I} + \sqrt{\frac{\pi}{2}}\, (1 + \nu_*) \tau_1 \right], \\
\av{u_{3}} (r) &= -4\sqrt{r}\, \frac{1 + \nu_*}{E_*}\, \tau_{3}, \quad \jump{u_{3}} (r) = 4 \sqrt{\frac{2r}{\pi}} \frac{1 + \nu_*}{E_*}\, K_{III}.
\end{aligned}
\end{equation}
It should be noted that by substituting the above equations into \eqref{err_final3a} one obtains the expression for
the Energy Release Rate \eqref{eq:generalERR} computed in Sect.~\ref{Sect:3}.
As such, the result presented above provides a resolution to the apparent disparity that motivated this work, demonstrating that updating the LEFM to incorporate singular tractions produces the ERR claimed in \citep{Wrobel2017}.

\section{Evaluation of the Energy Release Rate for special cases}\label{Sect:5}
\subsection{Comparison with ERR corresponding to the classical LEFM assumptions}\label{Sect:5a}

Before considering any application of the framework developed in Sect.~\ref{Sect:4}, we first show that it coincides with the classical LEFM if the singular tractions on the crack faces vanish, $\tau_1=\tau_2=\tau_3=0$.
In this case, Eq.\ \eqref{Sigma_Tang1} implies:
\[
 \jump{\sigma_{12}} (r) = \jump{\sigma_{r\theta}} (r) = 0, \quad
  \jump{\sigma_{22}} (r) = \jump{\sigma_{\theta\theta}} (r) =0,\quad
 \jump{\sigma_{23}} (r) = -\jump{\sigma_{\theta 3}} (r) = 0.
\]
Then, the ERR integrals \eqref{err_final2a} and \eqref{err_final3a} reduce to \eqref{err_final1}
or, equivalently, to
\[
\begin{aligned}
{\cal  G} =
\lim_{\Delta a \to 0} \frac{1}{2 \Delta a} \int_0^{\Delta a} \Bigg\{
& \av{\sigma_{\theta r}(a)(r)} \jump{-u_r(a)(\Delta a - r)} + \\
& \av{\sigma_{\theta \theta}(a)(r)} \jump{-u_\theta(a)(\Delta a - r)} + \av{\sigma_{\theta z}(a)(r)} \jump{u_z(a)(\Delta a - r)}
\Bigg\}\, dr.
\end{aligned}
\]
This expression coincides with the well-known representation of the crack closure integral or virtual crack closure integral \citep{Rybicki1977,Krueger2004}. Further, evaluating the ERR obtained in Sect.~\ref{Sect:3} in this case yields the formula \eqref{eq:generalERR}:
\begin{equation}
{\cal  G}= \frac{1}{E_*} \left( K_I^2 + K_{II}^2 + (1 + \nu_*) K_{III}^2 \right).
\end{equation}
Noting the definitions of $G_*$, $\nu_*$, in the plane strain regime this coincides with the classical Energy Release Rate, $J$, if the loading at the crack faces has a singularity weaker than the classical square root one.

\subsection{Hydraulic fracture accounting for shear traction induced on the crack surfaces}\label{Sect:5b}

The primary motivation for this work is a reconciliation of the ERR results obtained for hydraulic fracture with singular shear tractions at the crack faces with those from the classical LEFM. With the general formula for ERR now derived, we can check that it is consistent with the results from \citep{Wrobel2017} outlined in Sect.~\ref{Sect:2a}.

As the tangential traction in this case is only loaded by the fluid along the fracture walls, we consider the case when $\tau_2=\tau_3=0$. Then, the general form of the ERR \eqref{eq:generalERR} becomes
\begin{equation}
 \label{eq:generalERR_new0}
{\cal  G} = \frac{1}{E_*} \left( K_I^2 + K_{II}^2 + (1-\nu_*)\, K_{III}^2 + 2\sqrt{2\pi} \, K_I \tau_1 \right).
\end{equation}
which after using \eqref{tau_1} coincides with the energy release rate, $J$, computed in accordance with \eqref{ERR_definition} in
\eqref{err_fracture}, resolving the above-mentioned mismatch. Thus, both methods of computing the ERR - using Rice's formula \eqref{ERR_definition} and via the generalised Irwin's formula \eqref{err_final2a} (derived from the Principle of Virtual Work) - lead to the same result.

It may seem that \eqref{eq:generalERR_new0} implies that the shear stress induced by the fluid on the fracture walls should always promote the crack propagation (compare the signs of $K_I$, $\tau_1$). However, this is not universally true. Indeed, for HF incorporating tangential traction the Mode-I stress intensity factor, $K_I$, is defined differently than that from the classical case, $K_I^{cl}$. Thus, if $K_I$ decreases as a result of the acting shear stress ($K_I^{cl}>K_I$), then a direct conclusion can't be drawn and a separate analysis is required (see for example \citep{Perkowska2017}).

\subsection{Growth/shrinkage of a thin rigid 2D inclusion}\label{Sect:5c}

We note that the case of rigid inclusions (anticracks) embedded in an elastic medium, such as the cases discussed in Sect.~\ref{Sect:2ba}, require that the displacements on the different sides of the defect surfaces are equal. Thus, one can observe from the jumps of the displacements \eqref{Disp_Tang1}, that we must have
\[
 K_{I} + \sqrt{\frac{\pi}{2}}\,(1 + \nu_*)\tau_1=0,\quad   K_{II} - \sqrt{\frac{\pi}{2}}\,(1 + \nu_*) \tau_2=0.
\]

Indeed, if these relationships and the condition $\tau_3=0$ are substituted into the asymptotic representations \eqref{eq:asymu1} and \eqref{eq:asyms1}, one obtains the asymptotic representation evaluated by
\citep{Goudarzi2021}, \citep{Atkinson1995} (we use the plane strain notations for easy comparison) as $r\to0$:
\begin{equation}
\begin{bmatrix}
u_r \\
u_\theta
\end{bmatrix}
\sim \frac{1 + \nu}{E} \sqrt{\frac{r}{2 \pi}}
\Bigg\{
K_{I}
\begin{bmatrix}
\ds -\cos \frac{\theta}{2} \left[ 1 - (7 - 8\nu) \cos \theta \right] \\[3mm]
\ds -\frac{7 - 8\nu}{2} \csc \frac{\theta}{2} \sin^2 \theta
\end{bmatrix}
+ K_{II}
\begin{bmatrix}
\ds -\frac{5 - 8\nu}{2} \csc \frac{\theta}{2} \sin^2 \theta \\[3mm]
\ds -\cos \frac{\theta}{2} \left[ 1 + (5 - 8\nu) \cos \theta \right]
\end{bmatrix}
\Bigg\},
\end{equation}
\begin{equation}
\begin{bmatrix}
\sigma_{rr} \\
\sigma_{\theta\theta} \\
\sigma_{r\theta}
\end{bmatrix}
\sim \frac{1}{\sqrt{2\pi r}}
\Bigg\{
K_{I}
\begin{bmatrix}
\ds -\cos \frac{\theta}{2} \left[ \frac{1 - 4\nu}{2} - \frac{7 - 8\nu}{2} \cos \theta \right] \\[3mm]
\ds \cos \frac{\theta}{2} \left[ \frac{5 - 4\nu}{2} - \frac{7 - 8\nu}{2} \cos \theta \right] \\[3mm]
\ds -\sin \frac{\theta}{2} \left[ \frac{3 - 4\nu}{2} + \frac{7 - 8\nu}{2} \cos \theta \right]
\end{bmatrix}
+ K_{II}
\begin{bmatrix}
\ds -\sin \frac{\theta}{2} \left[ \frac{5 - 4\nu}{2} + \frac{5 - 8\nu}{2} \cos \theta \right] \\[3mm]
\ds \sin \frac{\theta}{2} \left[ \frac{1 - 4\nu}{2} + \frac{5 - 8\nu}{2} \cos \theta \right] \\[3mm]
\ds \cos \frac{\theta}{2} \left[ \frac{3 - 4\nu}{2} - \frac{5 - 8\nu}{2} \cos \theta \right]
\end{bmatrix}
\Bigg\}.
\end{equation}
Then, considering only deformation in $X\!OY$-plane ($K_{III}=0, \tau_3=0$) we obtain
\begin{equation}
\label{eq:rigid}
J ={\cal G}= -\frac{1 -\nu^2}{E}(3-4\nu) \left( K_I^2 + K_{II}^2 \right),
\end{equation}
which naturally coincides with the results of \citep{Atkinson1995,Goudarzi2021}. The ERR in this problem is always negative, the physical meaning (as noted by \citep{Goudarzi2021}) being that the reduction in the length of a particular inclusion would lead to a decrease in the total potential energy of the system. Note that the growth of the rigid inclusion is only possible by the action of additional (external) physical/chemical fields "freezing" the line in the front of the rigid inclusion. In such action, no mechanical energy is required; moreover, the energy will radiate from the inclusion into the body. Additionally, note that the `neutral orientation' of an inclusion - where it's presence does not affect the resulting stress field near the defect tip - can be obtained (e.g. the stress intensity factors $K_I=K_{II}=0$, see for details \citep{Goudarzi2021}).

\subsection{Closed crack under Mode I-II and a shear band}
\label{Sect:5d}
It is well known that in the LEFM framework open crack requires positive value of the SIF $K_I>0$ and if $K_I<0$ the classic solution leads to the crack surface overlap near the crack tip (see for example
\citep{Negative2021} and references therein).

Indeed, formally computed SIF under mode I is negative under compression and the crack is closed that eliminates the singularity. In the framework of the present analysis this fact can be easily shown.
Let's assume that, if one takes into account certain transmission conditions between the closed crack surfaces, it is possible to create such a stress-strain state that preserves the singular compressive stress ($\sigma_{\theta\theta}(r,0)<0$) on the crack line ahead the crack tip (as such the condition $K_I<0$ would be preserved). We discuss plane problem and thus $K_{III}=\tau_3=0$.
If the crack is closed and its surfaces are in contact (without presence of any external interference and regardless of the friction law), the following conditions are satisfied along the crack surfaces:
\[
[u_\theta]=0,\quad [\sigma_{r\theta}]=0,\quad [\sigma_{\theta\theta}]=0.
\]
According to \eqref{Sigma_Tang1} and \eqref{Disp_Tang1}, these three conditions are equivalent to the following expressions:
\[
\tau_1=0,\quad \tau_2=0,\quad  K_{I} + \sqrt{\frac{\pi}{2}}\, (1 + \nu_*) \tau_1 =0.
\]
Consequently, in the case of closed crack, the stress intensity factor $K_I$ is always equal to zero (and thus, cannot be negative) regardless of the friction law. The latter may affect only a value of the remaining SIF $K_{II}$ that can be positive, negative or zero depending on the given law and external load applied. Being completely different in its physical interpretation, the problem for a finite shear band appearing in a solid is mathematically equivalent to the analysis of the closed crack  \citep{Sonato2015,Shearband2015,Panos2021}.

\subsection{Growth of a Mode III crack in general settings}\label{Sect:5e}

Let us now consider the growth of a Mode III crack. The stress and displacements act in the z-direction only: $K_I=K_{II}=\tau_1=\tau_2=0$, with the stress and displacement fields given by \eqref{eq:asymu1aaa}, \eqref{eq:asyms1aaa}. The resulting ERR is obtained from \eqref{eq:generalERR}
\begin{equation}
\label{eq:rigid}
J={\cal G} = \frac{1 +\nu}{E}
\left( K_{III}^2  - 2\pi \tau_{3}^2 \right),
\end{equation}
and coincides with
\citep{Wang1986,Chaudhuri2012}. As a result, any additional asymmetric admissible (square root) singular shear traction applied to the crack surfaces near the crack tip always decreases the ERR in comparison with the classic Mode III crack (with only the symmetric load effecting the value of $K_{III}$). Note that in the case where the symmetric and asymmetric fields are coupled, this conclusion may be wrong and such case requires additional analysis, similarly to the above-discussed case of hydraulic fracture (see Sect.~\ref{Sect:5b}).

Note that locally, displacements resulting from the action of the stress $\pm\tau_3r^{-1/2}$ (see Table~\ref{tab:tractions}) at the crack surfaces always produce zero jump at the crack tip.
Thus, the corresponding solution is equivalent to that for a thin rigid half-plate extracted from the 3D body. Moreover, this case also corresponds to the Mode III crack with Gurtin-Murdoch surface stresses. Indeed, only symmetric (even) Mode-III square-root singular traction holds in such a case, as discussed in Sect.~\ref{Sect:2bb} (see \citep{Gorbushin2020} for more detail). In both cases, the value of the classic SIF $K_{III}$ in the formula \eqref{eq:rigid} appears to be equal to zero.

\section{Discussion and conclusions}\label{Sect:6}

The general form of the Energy Release Rate (ERR) accounting for tractions applied to the crack surfaces and having a singularity at the crack tip was obtained. The derivation was completed in two ways: from Rice's $J$-integral \eqref{eq:generalERR} and through the  principles of virtual work \eqref{err_final3a}. This involved updating the crack closure integral \eqref{e1}, with an additional term accounting for the action of the stress field ahead of the crack tip. The stress, displacement discontinuities and averages were given in terms of the stress intensity factors, $K_{j}$, and traction-induced stress intensity factors, $\tau_j$, ($j=1,2,3$) in  \eqref{Sigma_Tang1}--\eqref{Disp_Tang1}. Note that no assumptions were made concerning the relationships between the different constants (SIFs) participating in the analysis. This development constitutes a generalisation of the classical Linear Elastic Fracture Mechanics that accounts for the mentioned singular tractions. It is demonstrated that when the singular tractions on the crack faces were finite, the updated ERR \eqref{eq:generalERR} reduced to that of the classical LEFM.

This approach was compared with a number of examples from the literature in Sect.~\ref{Sect:5}. The updated formulation coincided with known results for hydraulic fracture (HF) with induced tangential traction on the fracture walls \citep{Wrobel2017}, the growth/shrinkage of 2D thin rigid inclusions within a soft domain \citep{Goudarzi2021}, and the growth/shrinkage of thin defects in Mode-III case.
The results offer a straightforward approach to determining the singular fields near the tip of a line-type defect in an elastic body, provided that boundary conditions on its surfaces are given (all assumptions are given in Sect.~\ref{Sect:4}).

For the case of HF with tangential traction on the crack walls, the apparent contradiction between the results of \citep{Wrobel2017} and formal application of the classic Irwin's crack closure integral has been resolved. The discrepancy was shown to be the consequence of not accounting for singular tractions effects. The resulting ERR \eqref{eq:generalERR} implies that the incorporation of the tangential traction may affect the fracture propagation, confirming the more detailed analysis given in \citep{Perkowska2017}.
Note that the updated ERR \eqref{err_final3a} was obtained assuming a propagating straight linear defect that preserves the same boundary conditions on its sides (see Sect.~\ref{Sect:4}). As a result, in more complex cases where the fracture topology is not preserved (crack redirection, or a mini crack initiation from the existing initial defect), additional analysis is needed to obtain the ERR for this specific configuration.
In such cases however, the results given here may be used as a benchmark for the simplified variant of the problem.

%


\section*{Acknowledgment}
The authors are grateful to Profs. D. Bigoni, D. Garagash, P. Papanastasiou, J.R. Rice and L.I. Slepyan for fruitful discussions and useful comments.
G.M. acknowledges Wolfson Research Merit Award from the Royal Society and Ser Cymru Future Generations Industrial Fellowship.

\section*{Funding}
A.P. gratefully acknowledges the funding from the European Union's Horizon 2020 research and innovation programme under the Marie Skłodowska-Curie grant agreement No 955944 - RE-FRACTURE2.
D.P. and G.M. would like to thank Ser Cymru II Research Programme by Welsh Government supported by European Regional Development Fund.
M.W. was supported by European Regional Development Fund and the Republic of Cyprus
through the Research Promotion Foundation (RESTART 2016 - 2020 PROGRAMMES, Excellence Hubs,
Project EXCELLENCE/1216/0481).

\section*{Declaration of Competing Interest}
The authors declare that there is no conflict of interest.

\bibliographystyle{plainnat}
\bibliography{Bibliography}

\end{document}